\documentclass[aps,prl,showpacs,twocolumn,superscriptaddress]{revtex4}
\usepackage{graphicx}
\usepackage{bm}
\usepackage{amsmath}

\def\Im{\rm{Im}}
\def\ra{\rangle}
\def\la{\langle}
\def\be{\begin{equation}}
\def\ee{\end{equation}}
\def\bea{\begin{eqnarray}}
\def\eea{\end{eqnarray}}

\newcommand{\Sin}{{\rm sin}}

\begin{document}
\title{Dynamical magneto-electric coupling in helical magnets}

\author{Hosho Katsura}
\email{katsura@appi.t.u-tokyo.ac.jp}
\affiliation{Department of Applied Physics, the University of Tokyo,
7-3-1, Hongo, Bunkyo-ku, Tokyo 113-8656, Japan}

\author{Alexander V. Balatsky}
\email[Electronic address: ]{avb@lanl.gov} \affiliation{Theoretical
Division, Los Alamos National  Laboratory, Los Alamos, New Mexico
87545 USA}

\author{Naoto Nagaosa}
\email{nagaosa@appi.t.u-tokyo.ac.jp}
\affiliation{Department of Applied Physics, the University of Tokyo,
7-3-1, Hongo, Bunkyo-ku, Tokyo 113-8656, Japan}
\affiliation{Correlated Electron Research Center (CERC),
National Institute of Advanced Industrial Science and Technology (AIST),
Tsukuba Central 4, Tsukuba 305-8562, Japan}
\affiliation{CREST, Japan Science and Technology Agency (JST), Japan}

\begin{abstract}
Collective mode dynamics of the helical magnets coupled to electric 
polarization via spin-orbit interaction is studied theoretically. 
The soft modes associated with the ferroelectricity are not the 
transverse optical phonons, as expected from the Lyddane-Sachs-Teller 
relation, but are the spin waves hybridized with the electric 
polarization. This leads to the Drude-like dielectric function 
$\varepsilon(\omega)$ in the limit of zero magnetic anisotropy. 
There are two more low-lying modes; phason of the spiral and rotation of 
helical plane along the polarization axis. The roles of these soft 
modes in the neutron scattering and antiferromagnetic resonance are 
revealed, and a novel experiment to detect the dynamical 
magneto-electric coupling is proposed. 
\end{abstract}

\maketitle

The gigantic coupling between the 
magnetism and ferroelectric properties is
now an issue of keen interest
\cite{KimuraNature,KimuraPRB,Goto,Arima,Kenzelmann,Lawes,Chapon,Blake, 
KimuraPRL}. 
A representative system of interest is $R$MnO$_3$ with $R$=Gd,Tb,Dy. 
For example, TbMnO$_3$ shows 
a ferroelectric moment $P//c$ below a temperature
$T_{FE}=28K(<T_{Neel}=42)$, and furthermore changes its direction 
of electric polarization to $a$ axis under the magnetic field $H//b$ \cite{KimuraNature,KimuraPRB,Goto,Arima}.
Similar strong coupling behavior has been observed in Ni$_3$V$_2$O$_8$
\cite{Lawes}, $R$Mn$_2$O$_5$ ($R$=Tb,Ho,Dy)\cite{Chapon,Blake}, 
Ba$_{0.5}$Sr$_{1.5}$Zn$_2$Fe$_{12}$O$_{22}$ \cite{KimuraPRL}.
A common and essential feature of these compounds is that there are
frustrations in the magnetic interactions. For $R$MnO$_3$, 
Kimura {\it et al}. 
\cite{KimuraPRB} revealed that the increased GdFeO$_3$-type distortion 
of perovskite lattice leads to the further-neighbor exchange interactions
and nontrivial magnetic structures. Later the neutron scattering experiment
determined the spin structure of TbMnO$_3$; it shows the incommensurate 
collinear spin ordering pointing along $b$-direction for
$T_{FE}<T<T_{Neel}$, while the helical spin structure winding within the
$b-c$ plane occurs for $T<T_{FE}$ with the helical wave vector 
$q//b$ \cite{Kenzelmann}. 
A similar helical structure is also observed in
Ni$_3$V$_2$O$_8$ \cite{Lawes}. 
These experiments point to key role
of the non-collinear spin configurations such as helical spin structure, which are
induced by frustrated exchange interactions, in producing the electric polarization
and enhanced magneto-electric coupling.

A microscopic mechanism of the ferroelectricity of 
magnetic origin 
has recently been proposed by Katsura {\it et al}. \cite{Katsura}, which 
is based on the idea that spin current is induced between the noncolliear 
spins and hence is the electric moment due to Aharonov-Casher 
effect \cite{AC,AB}. 
This result can be regarded as an inverse effect of Dzyaloshinskii-Moriya 
interaction 
\cite{DM}. Phenomenological treatment of
this mechanism \cite{Mostovoy} and its extension to include
electron-lattice interaction \cite{Dagotto} have been also reported.
By now the static and/or ground state properties of the magneto-electric
coupled systems are well understood. The obvious next step is
to study their novel dynamic properties searching for the unique 
electromagnetic
effects. In this context, we recall that the collective modes are a central issue 
in the research on ferroelectrics, where one of the transverse optical (TO) 
phonons softens toward the 
phase transition and is condensed according to the Lyddane-Sachs-Teller 
relation.
This conventional view is not relevant in the case of multiferroics since the 
lattice displacement is not essential to the electronic polarization. 
Therefore it is
an important open problem to identify relevant collective mode 
that are responsible for ferroelectricity of magnetic origin. 

In this paper, we study exactly this question, i.e., the low energy dynamics of the system of the coupled
spins and electric polarization. We found the new
collective modes of spin and polarization waves
that couple the 
dielectric and magnetic properties in a novel way. 
Experiments aimed at 
the detection of these
collective modes in terms of the magnetic resonance, neutron scattering
and the ac-dielectric measurement are also discussed. 

\begin{figure}
\includegraphics[width=8cm,clip]{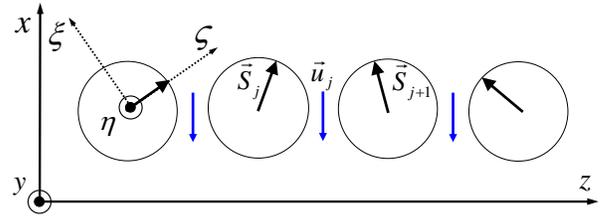}
\caption{Schematic ground state configuration of spins(black arrows) and 
lattice displacements(blue arrows).}
\end{figure}
In ref. \cite{Katsura}, Katsura {\it et al}. have shown that spin 
(super)current in noncollinear magnets 
$\vec j_s(\propto \vec S_i \times \vec S_j)$ leads 
to the electric polarization $\vec P \propto \vec e_{ij} \times \vec j_s$ 
with $\vec e_{ij}$ being the unit vector connecting two sites $i$ and $j$. 
An effective Hamiltonian describing this
coupling between the spins and the atomic displacement $\vec u_i$,
which represents the electric polarization including the displacement of electronic cloud,
is given as;
\begin{eqnarray}
H&=&H_1 + H_2 + H_3+H_4,
\\
H_1 &=& -\sum_{m,n} J(R_m-R_n) \vec S_m \cdot \vec S_n,
\\
H_2 &=& -\lambda \sum_m (\vec u_m \times \vec e_z)\cdot(\vec S_m \times
\vec
S_{m+1}),
\\
H_3 &=& \sum_m (\frac{\kappa}{2}{\vec u_m}^2+\frac{1}{2M} {\vec P_m}^2),
\\
H_4&=& \sum_m D(S^y_m)^2.
\end{eqnarray}
Here, the spin-lattice interaction $\lambda$ stems from the relativistic spin-orbit
interaction and corresponds to the DM interaction once the static displacement 
$\langle{\vec u}_j\rangle$ is non-zero 
and breaks the inversion symmetry. Note that purely electronic polarization 
is possible even without the atomic displacement, but the latter is always 
accompanied with the inversion symmetry breaking. 
Therefore ${\vec u_m}$ should be regarded as 
the lowest frequency representative coordinates relevant to the polarization, i.e., 
the transverse optical phonon, and the polarization $\vec p$ is given by $e^*{\vec u_m}$ with a Born 
charge $e^*$. The term $D(S^y)^2$ with positive $D$ represents the easy-plane
spin anisotropy. Here, we have assumed that the ground state spin 
configuration on 
the plane perpendicular to the helical wave vector is ferromagnetic and the 
low-energy excitations on this plane have a usual parabolic dispersion 
relation, i.e., $\propto |{\vec q_{\perp}}|^2$. Hence we shall focus on the 
modes along the helical wave vector which is set parallel to the $z$-axis 
shown in Fig.1. 
In $H_1$, the summation is taken over all the combinations of $m,n$ and 
$J(R_m-R_n)$ denotes the Heisenberg interaction between $\vec S_m$ and 
$\vec 
S_n$ at $R_m$ and $R_n$, respectively. 
First, we determine the classical ground state configuration of spins and 
lattice distortions. Since we consider the static configuration, the 
kinetic 
term in $H_3$ can be negligible. The variation of $H$ with respect to 
$\vec u_m$ gives 
$\vec u_m = \frac{\lambda}{\kappa} \vec e_z \times (\vec S_m \times \vec 
S_{m+1}).$
Henceforth we assume that spins are on the easy plane, i.e., $zx$-plane 
and explicitly given by
$S_n^z = S {\rm cos}(QR_n + \phi)$, 
$S_n^x = S {\rm sin}(QR_n + \phi)$,
$S_n^y = 0$.
By substituting this spin configuration into the energy
$\varepsilon$ per spin is rewritten as
$ \varepsilon = -S^2 {\tilde J(Q)} $ 
with $\tilde J(q)\equiv J(q)+\frac{\lambda^2 S^2}{2\kappa}{\rm sin}^2(qa)$.
Here $J(q) \equiv \sum_m J(R_m)e^{-iqR_m}$ is the Fourier transform of 
$J(R_m)$ 
and $a$ denotes the lattice constant. The wave number $Q$ is 
determined to maximize $\tilde J(q)$. 
Using $Q$, the uniform lattice 
displacement 
is given as
$
\vec u_m = -\frac{\lambda S^2}{\kappa} {\rm sin}(Qa)\vec e_x,
$
and hence the static electric polarization is given as 
${\vec p}=-p{\vec e_x}$ with $p=e^* u^x_m=e^* \frac{\lambda S^2}{\kappa}{\rm sin}(Qa)$.

Now we consider the equations of motion for spins and displacements, and 
study the collective modes of our system. 
We introduce a rotating local coordinate system $\xi, \eta,\zeta$ 
such that the $\zeta$-axis coincides with the equilibrium spin direction 
at each site, the $\xi$-axis is perpendicular to this direction in 
$zx$-plane 
and the $\eta$-axis parallel to the $y$-axis (see Fig.1) \cite{Nagamiya}.
Assuming the small quantum/thermal fluctuations around the classical configuration,
$S^{\zeta}_n$ and $u^x_n$ can be regarded as a 
constant and the equation of motion can be written as
$\dot S^{\xi}_n = S^{\eta}_n H^{\zeta}_n - S H^{\eta}_n$,
$\dot S^{\eta}_n = S H^{\xi}_n - S^{\xi}_n H^{\zeta}_n$, 
where $S=S^{\zeta}_n=$const., and ${\vec 
H_n}=(H^{\xi}_n,H^{\eta}_n,H^{\zeta}_n)$ 
is the effective field acting on the $n$-th spin explicitly
given by
\begin{equation}
\left\{
\begin{array}{l}
H^{\zeta}_n = 2S\left(J(Q)+\dfrac{\lambda^2 S^2}{\kappa}{\rm 
sin}^2(Qa)\right)\vspace{0.5em}\\
H^{\xi}_n = \displaystyle\sum_m 2J(R_m-R_n)S^{\xi}_m {\rm 
cos}(Q(R_m-R_n))\vspace{0.3em}\\
+ \dfrac{\lambda^2 S^2}{\kappa} {\rm 
sin}^2(Qa)(S^{\xi}_{n+1}+S^{\xi}_{n-1})\vspace{0.5em}\\
H^{\eta}_n = \displaystyle\sum_m 2J(R_m-R_n)S^{\eta}_m -2DS^{\eta}_n 
\vspace{0.3em}\\
+ \lambda S\left(u^{y}_n{\rm cos}(QR_{n+1}+\phi)-u^{y}_{n-1}{\rm 
cos}(QR_{n-1}+\phi)\right).
\end{array}
\right.
\end{equation}
Here, we regard $S^{\xi},S^{\eta}$ and $u^y$ as sufficiently small and 
neglect their second- or higher order terms. We can also write down the 
equation of motion for $u^y$.
Introducing the Fourier components of 
$S^{\xi}_m,S^{\eta}_m,u^y_m$ and $p^y_m$,
the equations of motion are given by 
\begin{equation}
\begin{array}{l}
\dot S^{\eta}_q = -A(q)S^{\xi}_q, \vspace{0.5em}\\
\dot S^{\xi}_q = B(q)S^{\eta}_q
-i\lambda S^2 \left(
\dfrac{e^{iQa}-e^{-iqa}}{2i}e^{i\phi} u_{q-Q}+
\Big(
\begin{array}{c}
Q \to -Q \\
\phi \to -\phi\\
\end{array} \Big)
\right),\vspace{0.5em}\\
\dot u_q = \dfrac{1}{M}p_q, \vspace{0.5em}\\
\dot p_q = -\kappa u_q 
+i\lambda S \left(
\dfrac{e^{iQa}-e^{i(q-Q)a}}{2i}e^{i\phi} S^{\eta}_{q-Q} 
+\Big(\begin{array}{c}
Q \to -Q\\
\phi \to -\phi
\end{array}\Big)\right),\\
\end{array}
\label{eomk}
\end{equation}
where
\begin{eqnarray}
A(q) &=& 2S \bigg[
J(Q)-\frac{J(Q+q)+J(Q-q)}{2} \nonumber 
\\ 
& &+\frac{2\lambda^2 S^2}{\kappa} {\rm sin}^2(qa/2){\rm sin}^2(Qa)
\bigg],
\nonumber \\
B(q) &=& 2S\left[J(Q)-J(q)+\frac{\lambda^2 S^2}{\kappa}{\rm 
sin}^2(Qa)+D\right].
\end{eqnarray} 
We note that $A(q)$ and $B(q)$ satisfy the relations 
$A(-q)=A(q),B(-q)=B(q)$, respectively, and $A(0)$ is equal to zero.

From Eq.(\ref{eomk}), one can see the coupling between the spin wave modes 
and electric polarization. First $S^\eta$ and $S^\xi$ are canonical 
conjugate variables, and form a harmonic oscillator at each $q$ in the 
rotated frame. This spin wave at $q$ is coupled with the 
phonon
$u$ at $q \pm Q$, or $u_q$ is coupled to $S^\eta/S^\xi$ at $q \pm Q$. 
Especially the uniform lattice displacement $u^y_{0}$ is coupled to 
$e^{-i\phi}S^{\alpha}_Q -e^{i\phi} S^{\alpha}_{-Q}$ $(\alpha=\eta,\xi)$,
which corresponds to the rotation of both the spin plane and the 
direction of the 
polarization along the $z$-axis. 
This mode is the Goldstone boson with frequency $\omega=0$ 
when $D=0$, i.e., the symmetric case around $z$-axis.
On the other hand, $e^{-i\phi}S^{\alpha}_Q + e^{i\phi}S^{\alpha}_{-Q}$ 
corresponds to the 
rotation of spin plane along $x$-axis, which is decoupled to the 
polarization but is gapped by the effective spin anisotropy introduced by the
spin-lattice interaction.
The spin wave mode at $q=0$ corresponds to the sliding mode, i.e., phason,
of the spiral. This is decoupled from $u$ and has zero energy at 
$q=0$.

Using Eq.(\ref{eomk}) with canonical commutation relations 
$[u_q, p_{q'}]=i\delta_{q,-q'}$ and $[ S^{\xi}_q, 
S^{\eta}_{q'}]=iS \delta_{q,-q'}$, the matrix form of the retarded 
Green's function 
\begin{eqnarray}
G^R(AB;t-t') &\equiv& -i \theta(t-t') \langle [A(t),B(t')]\rangle,
\nonumber \\
G^R(AB;\omega) &\equiv& \frac{1}{2\pi} \int^{\infty}_{-\infty} G^R_{AB}(t) 
e^{i\omega t},
\label{Green}
\end{eqnarray}
($A,B = u_q, p_q, S^{\xi}_q,S^{\eta}_q$) is obtained. The imaginary part of 
the ac susceptibility is also defined as 
$\chi''(AB;\omega)\equiv -{\rm Im}G^R(AB;\omega)$. 
To be more explicit, we will discuss below some physically important cases.
\vskip 0.3cm
\noindent
{\it ac dielectric properties}\hspace{1mm}-- The dynamical dielectric 
function is given by 
$\varepsilon_{yy}(\omega) = 1 + 4 \pi (e^*)^2 G^R(u_0 u_0; \omega)$,
where the $e^*$ is the Born charge corresponding to the displacement $u^y$ 
along $y$-axis, and 
\begin{equation}
G^R(u_0u_0;\omega)=\frac{\omega^2-\omega^2_p}
{2\pi M (\omega^4-(\omega^2_0+\omega^2_p)\omega^2+A(Q)D\omega^2_0)},
\end{equation}
where $\omega_p=\sqrt{A(Q) B(Q)}$
is the frequency of the spin-plane rotation mode along $x$-axis, i.e.,
$e^{-i\phi}S^{\alpha}_Q + 
e^{i\phi}S^{\alpha}_{-Q}$, and $\omega_0=\sqrt{\kappa/M}$ is that for the 
original phonon. This response function has the poles at
$\omega_{\pm}$, which are explicitly given by
\begin{equation}
\omega^2_{\pm} = \frac{1}{2}
\left(\omega^2_0+\omega^2_p
\pm \sqrt{(\omega^2_0+\omega^2_p)^2-4A(Q)D\omega^2_0}
\right)
\end{equation}
Assuming $D, \lambda \ll \omega^2_0$, one can estimate as
$\omega_- \cong \sqrt{ A(Q) D} \sim \sqrt{8SJD}$, and
$\omega_+ \cong \omega_0$.
With this frequencies, one can write 
$\varepsilon_{yy}(\omega) = 1 + \sum_{\pm}\omega_{\pm} I_{\pm}/(\omega^2 -
\omega_\pm^2)$ with the ``oscillator strengths'' $I_\pm$ 
being given by 
$I_- =\dfrac{2 (e^*)^2 (\omega^2_p-\omega^2_-)}{ M \omega_- (\omega^2_+ 
-\omega^2_-)}$, and 
$I_+ = \dfrac{2 (e^*)^2 ( \omega^2_+ - \omega^2_p)}
{M \omega_+ (\omega^2_+ -\omega^2_-)}$. Note that the integral 
$ - \int_0^{\infty}d\omega \varepsilon_{yy}(\omega)$ is given by 
$(\pi/2) ( I_- + I_+ )$. 
Here, the two modes contributing to the dielectric function are (i)
the phonon mode with the frequency $\omega_+ \cong \omega_0$,
and (ii) the $z$-axis rotation mode at $\omega_- \cong \sqrt{A(Q)D}$. 
Usually 
$\omega_0$ is a high frequency and does not show any softening 
in the present case.
The ferroelectricity is due to the spin ordering, which is hybridized 
with the polarization mode. In other words, the low frequency dielectric
function is mostly due to the spin wave mode at $\pm Q$. Note the 
low-frequency
behavior of $\varepsilon_{yy}(\omega)$ is similar to the Drude form when
$D=0$, and the oscillator strength $I_-$ (at $\omega=\omega_-$)
is enhanced as $ I_- \sim p^2 \sqrt{J/D}$ as $ D \to 0$.
This means that even though the spin-polarization coupling $\lambda$ and hence the static polarization $p$ is small,
the spin wave mode can contribute significantly when $D$ is small.
\vskip 0.3cm
\noindent
{\it neutron scattering spectra}\hspace{1mm} --
Next, we shall examine how the spin-polarization coupling affects 
the (inelastic) neutron scattering spectra. 
The intensity of the neutron scattered with
the momentum transfer $q$ is estimated as

$
\frac{d^2 \sigma}{d\Omega d\omega} \propto
- {\Im} \sum_{\alpha \in \perp} G^R(S^{\alpha}_q S^{\alpha}_{-q};\omega),
$
where the superscript $\alpha$ indicates the components perpendicular to 
$\vec q$. 
In what follows, for the sake of simplicity, we set $D=0, \phi=0$. 
In order to study the neutron scattering spectra, it is needed to 
rotate the spin coordinates back to the original laboratory system
$S^a$ ($a=x,y,z$). We can easily derive the transformation formula given by
\begin{equation}
\left\{
\begin{array}{l}
G^R(S^x_qS^x_{-q};\omega)=\dfrac{1}{4}
\displaystyle\sum_{p=\pm Q \atop p'=\pm 
Q}G^R(S^{\xi}_{q+p}S^{\xi}_{-q-p'};\omega),\vspace{0.4em}\\
G^R(S^y_qS^y_{-q};\omega)=G^R(S^{\eta}_qS^{\eta}_{-q};\omega),\vspace{0.4em}\\
G^R(S^z_qS^z_{-q};\omega)=\dfrac{1}{4}\displaystyle\sum_{p=\pm Q \atop 
p'=\pm Q}
{\rm sgn}(pp')G^R(S^{\xi}_{q+p}S^{\xi}_{-q-p'};\omega).
\end{array}
\right.
\label{transf}
\end{equation}
For example, 
at the helical wavevector
${\vec q}=(0,0,Q)$, 
one can read the neutron scattering spectra
by using the above transformation formula (\ref{transf})
and the list of pole positions 
and corresponding intensities summarized in Table \ref{neutron}(a) and (b).
Note that the Green's function is written as
$G^R(\omega) = \sum_i a_i/(\omega^2-\omega_i^2)$  using the intensity $a_i$ and 
the pole position $\omega_i$. 
Now 4 modes are expected to contribute to the neutron spectrum \cite{comment};
(i) phason mode at $\omega=0$, (ii) $x$-axis rotation mode at 
$\omega= \omega_p$, (iii) phonon mode with $q=0$ at 
$\omega=\sqrt{\omega_0^2+ \omega_p^2}$, and (iv) phonon mode with
$q=Q$ at $\omega_0$. The former two, i.e., (i) and (ii), are of magnetic 
origin, and can be detected without the spin-polarization coupling, 
while the weights of the latter two, i.e., (iii) and (iv), are 
borrowed from the spin wave, and hence their relative magnitudes 
to the former ones are roughly estimated as; 
(iii)/(i)$ \sim (\omega_p/\omega_0)^2$, and
(iv)/(i) $ \sim \lambda^2/(\kappa J)$.

\begin{table}
(a)$G^R(S^y_QS^y_{-Q};\omega)$\\
\begin{tabular}{|l|c|c|c|}
\hline
$\omega_i$ & 0 & $\omega_p$ & $\sqrt{\omega^2_0+\omega^2_p}$\\
\hline
$a_i$ & $\dfrac{SA(Q)}{4 \pi} \dfrac{\omega^2_0}{\omega^2_0+\omega^2_p}$ 
& $\dfrac{SA(Q)}{4 \pi}$ & $\dfrac{SA(Q)}{4 \pi} 
\dfrac{\omega^2_p}{\omega^2_0+\omega^2_p}$\\
\hline 
\end{tabular}
\vspace{1.5mm}
\\
(b)$G^R(S^x_QS^x_{-Q};\omega)=G^R(S^z_QS^z_{-Q};\omega)$\\
\begin{tabular}{|l|c|c|}
\hline
$\omega_i$ & 0 & $\omega_0$\\
\hline
$a_i$ & $\dfrac{S}{8\pi}\left(B(0)-\dfrac{2 \lambda^2 S^3}{\kappa}{\rm 
sin}^2\left(\dfrac{Qa}{2}\right)\right)$ 
& $\dfrac{S}{8\pi}\dfrac{2 \lambda^2 S^3}{\kappa}{\rm 
sin}^2\left(\dfrac{Qa}{2}\right)$\\
\hline
\end{tabular}
\caption{The pole positions $\omega_i$ and corresponding intensities $a_i$
for each Green's function.}
\label{neutron}
\end{table}
\vskip 0.3cm
\noindent
{\it antiferromagnetic resonance}\hspace{1mm}--
We shall briefly discuss the antiferromagnetic resonance in our 
system. Here, $D=0$, and $\phi=0$ are also assumed.
Magnetic resonance experiments pick up uniform magnetic excitations and 
corresponding Green's functions are $G^R(S^a_0 S^a_0;\omega),
\hspace{3mm}(a=x,y,z)$, where axis $a$ corresponds to the 
direction of applying oscillating magnetic field.
Using again the transformation formula (\ref{transf}), 
one can see the pole of 
$G^R(S^x_0S^x_0;\omega)$ occurs at 
$\omega=\omega_p$ with the intensity $SB(Q)/(4 \pi)$, while
that of 
$G^R(S^z_0S^z_0;\omega)$ at 
$\omega = \sqrt{\omega_0^2 + \omega_p^2}$ with the intensity
$SB(Q)/(4 \pi)$. Note here that there is no dynamics of
$S^y_0$ since it is a conserved quantity.
The phonon mode with $q=\pm Q$ has a high frequency
$\omega= \sqrt{\omega_0^2+\omega_p^2}$ and should be difficult 
to be observed experimentally. The $x$-axis rotation mode
of low frequency ($\omega= \omega_p$) is not coupled directly
to the polarization, while the resonance intensity is due to 
the spin anisotropy induced by the spin-lattice coupling $\lambda$
and/or the spin anisotropy $D$ as discussed earlier \cite{Foner}. 

Now we turn to the realistic estimation of the parameters
in our theory extracted from the experiments on 
$R$MnO$_3$. 
The exchange coupling constant can be estimated from the 
spin wave dispersion observed in the neutron scattering 
experiment \cite{Kajimoto}. They estimated the exchange coupling
$J_1$ between the nearest neighbor spins as 
$8S J_1 \cong 9$ meV for PrMnO$_3$ and 
$8S J_1 \cong 2.4$ meV for TbMnO$_3$.
On the other hand, the anisotropy energy, which corresponds 
to $D$ in our theory, is around 0.4meV for both compounds. 
From the ESR spectra of Jahn-Teller distorted 
LaMnO$_{3+\delta}$($0\leq\delta\leq0.07$), Dzyaloshinskii-Moriya coupling 
is estimated as ${\cal D}\sim 0.1$meV \cite{Tovar} and hence we can 
evaluate the spin-lattice coupling $\lambda \sim 1$meV/{\rm \AA} 
\cite{Dagotto}. 
Assuming $\kappa \sim 1$eV/${\rm \AA}^2$ \cite{Dagotto}, we can obtain the 
static 
displacement $u_m=10^{-3}{\rm \AA}$. 
Another important quantity is the Born charge $e^*$.
This can be estimated from the value of electric polarization 
with the above $u_m$ and the lattice constant $a\sim 5{\rm \AA}$.
The electric polarization along $c$-axis is determined as 
$P_c \sim 0.2\mu$C/cm$^2$ for DyMnO$_3$ \cite{Goto} and hence 
we can estimate the Born charge $e^*$ as $16e$ where $e$ is 
the bare unit charge. 
This result, i.e., the Born charge is much greater than the unit charge, 
indicates that the electric polarization mainly consists of the 
displacement of the electronic cloud. 
At this moment, let us check the consistency of our estimation with 
another recent experiment on dielectric response \cite{Pimenov}. 
They observed the peak of $\Im \varepsilon$ at around
20cm$^{-1}$ with the magnitude of $1 \sim 2$ in 
GdMnO$_3$ and TbMnO$_3$. This 20cm$^{-1}$ is identified
with $\omega_-$ in Eq.(12), and the integration of 
$- {\rm Im}\varepsilon_{yy}(\omega)$ over 
$\omega$ gives the "oscillator strength" $(\pi/2)I_-$. 
From their results, we can estimate $I_- \sim 12$cm$^{-1}$. 
On the other hand, $\omega_-$ and $I_-$ are independently estimated as $\omega_-\sim 10$cm$^{-1}$ and $I_-\sim$4cm$^{-1}$ 
from the parameters determined above. 
Here, we have assumed that $S=2$ and $Qa=\pi/4$.
Both $\omega_-$'s and $I_-$'s,
are of the same order and hence 
we can conclude that our phenomenological model well describes the 
magneto-electric coupling in $R$MnO$_3$.
It turned out that the coupling between the 
spin wave and the transverse phonon is weak,
and is characterized by the ratio $\lambda^2/(\kappa J) \sim 10^{-3}$.
Therefore, it would be rather difficult to see this coupling in 
the neutron scattering experiment, i.e., the magnetic scattering
by the phonon modes. However, note that the low energy
dielectric response reflects the spin wave mode as the
Goldstone boson, i.e., $z$-axis spin rotation mode, and
it would be enhanced more if the spin anisotropy is weaker.

Lastly, we propose a novel experiment to see directly the 
dynamical magneto-electric coupling. Namely we can excite the
spin wave by electric field. These responses are described by the 
off-diagonal element of the Green's function;    
\begin{eqnarray}
&&G^R (u^y_0 S^z_0;\omega)= -G^R(S^z_0 u^y_0;\omega)
\nonumber \\
&&=\frac{1}{2\pi M}
\frac{i \lambda S^2\Sin(Qa)\omega}{\omega^4-(\omega^2_0 
+\omega^2_p)\omega^2+A(Q) D\omega^2_0}.
\end{eqnarray}
Proposed experiment requires to put the sample between the two electrodes and apply the a.c. 
electric field $E_y$ of frequency $\omega$
along $y$-direction. Then the magnetization of the same frequency 
$\omega$ can be detected at the edges of the sample 
polarized along $z$-direction. Now $\omega_- \cong 20$cm$^{-1}$ 
corresponds to the Tera-Hertz region, and the highest intensity
of the electric field there is $\sim 10^{3}$V/cm. 
Putting this value and all the other parameters estimated above
with overdamping $\gamma \sim \omega_-=20$cm$^{-1}$,
we obtain the estimation for the magnitude of the a.c. 
magnetization as $m^z \sim 0.4\times 10^{-4} \mu_{\rm B}$/Mn atom, which 
can be detected by Kerr rotation spectroscopy.
We should note here that the magnitude of $m^z$ is much enhanced
when $\gamma$ is small.
\noindent

The authors are grateful to
T. Arima, Y. Tokura, R. Kajimoto, T. Kimura and T. Lookman 
for fruitful discussions.
This work is financially supported by NAREGI Grant, Grant-in-Aids
from the Ministry of Education, Culture, Sports, Science and Technology
of Japan and by the US Department of Energy.


\begin{thebibliography}{99}

\bibitem{KimuraNature} 
T. Kimura et al., Nature \textbf{426}, 55 (2003).
\bibitem{KimuraPRB} 
T. Kimura et al., Phys. Rev. B\textbf{68}, 060403(R) (2003).
\bibitem{Goto} 
T. Goto et al., Phys. Rev. Lett.\textbf{92}, 257201 (2004).
\bibitem{Arima} 
T. Arima et al., Phys. Rev. B.\textbf{72}, 100102 (2005).
\bibitem{Lawes} 
G. Lawes et al., Phys. Rev. Lett.\textbf{95}, 087205 (2005).
\bibitem{Chapon} 
L. C. Chapon et al., Phys. Rev. Lett.\textbf{93}, 177402 (2004).
\bibitem{Blake} 
G. R. Blake et al., Phys. Rev. B.\textbf{71}, 177402 (2004).
\bibitem{KimuraPRL} T. Kimura, G. Lawes , and A. P. Ramirez,
Phys. Rev. Lett.\textbf{94}, 137201 (2005).
\bibitem{Kenzelmann} 
M. Kenzelmann et al.,Phys. Rev. Lett.\textbf{95},087206 (2005).
\bibitem{Katsura} 
H. Katsura, N. Nagaosa, and A.V.Balatsky, Phys. Rev. Lett. 
\textbf{95},057205 (2005).
\bibitem{AC} 
Y. Aharonov and A. Casher,Phys. Rev. Lett.\textbf{53}, 319(1984). 
\bibitem{AB}
A. V. Balatsky and B. L. Altshuler, Phys. Rev. Lett. \textbf{70}, 1678 
(1993).
\bibitem{DM} 
I. Dzyaloshinskii, J. Phys. Chem. Solids \textbf{4}, 241 (1958); 
T. Moriya, Phys. Rev. \textbf{120}, 91 (1960). 
\bibitem{Mostovoy} M. Mostovoy, cond-mat/0510692.
\bibitem{Dagotto} I.A. Sergienko, E. Dagotto, cond-mat/0508075.
\bibitem{Nagamiya}
T. Nagamiya, in {\it Solid State Physics}, edited by F. Seitz, D. Turnbull, 


and H. Ehrenreich (Academic Press, New York, 1967), Vol.20, p.305.

\bibitem{comment}
We should note here that we have neglected high-frequency correlations 
such as $\la S_{2Q}S_0\ra$, since the modes $S^{\alpha}_{\pm 2Q}$ have 
an energy higher than that of $S^{\alpha}_0$ or $S^{\alpha}_{\pm Q}$.

\bibitem{Foner}
S. Foner, \textit{Magnetism I}, edited by G. T. Rado and H. Suhl, (Academic Press, New York and London, 1963) p.383. 

\bibitem{Kajimoto}
R. Kajimoto {\it et al}, 
J. Phys. Soc. Jpn. \textbf{74}, 2430 (2005).

\bibitem{Tovar}
M. Tovar {\it et al.}, Phys. Rev. B{\bf 60}, 10199 (1999).

\bibitem{Pimenov}
A. Pimenov {\it et al}.,
to appear in Nature Physics(cond-mat/0602173). 


\end{thebibliography}
\end{document}